\documentclass[referee, linelo ,pdflatex,sn-nature]{sn-jnl}%

\usepackage{graphicx}%
\usepackage{multirow}%
\usepackage{amsmath,amssymb,amsfonts}%
\usepackage{amsthm}%
\usepackage{mathrsfs}%
\usepackage[title]{appendix}%
\usepackage{xcolor}%
\usepackage{textcomp}%
\usepackage{manyfoot}%
\usepackage{booktabs}%
\usepackage{algorithm}%
\usepackage{algorithmicx}%
\usepackage{algpseudocode}%
\usepackage{listings}%
\usepackage{epstopdf}
\usepackage[table]{xcolor}
\usepackage{hyperref} %

\definecolor{bossbg}{RGB}{230,245,255}       %
\definecolor{mdsurfbg}{RGB}{235,250,235}     %
\definecolor{mdmetalbg}{RGB}{255,240,225}    %
\definecolor{aissbg}{RGB}{245,235,255}       %
\definecolor{nmsbg}{RGB}{240,240,240}        %

\usepackage[version=4]{mhchem}

\title[Article Title]
{MAD-SURF: a machine learning interatomic potential for molecular adsorption on coinage metal surfaces
}         

\author[1]{\fnm{Manuel} \sur{Gonz\'alez Lastre}}
\equalcont{These authors contributed equally to this work.}

\author*[2,3]{\fnm{Joakim S.} \sur{Jestil\"a}}
\equalcont{These authors contributed equally to this work.}
\email{joakim.jestila@aalto.fi}

\author[1,4]{\fnm{Rub\'en} \sur{P\'erez}}

\author*[2,5]{\fnm{Adam S.} \sur{Foster}}
\email{adam.foster@aalto.fi
}
\affil[1]{\orgdiv{Departamento de F\'isica Te\'orica de la Materia Condensada}, 
\orgname{Universidad Aut\'onoma de Madrid}, \orgaddress{\postcode{28049}, \city{Madrid}, \country{Spain}}}

\affil[2]{\orgdiv{Department of Applied Physics}, \orgname{Aalto University}, 
\orgaddress{\postcode{00076}, \city{Espoo}, \country{Finland}}}

\affil[3]{\orgdiv{Department of Chemistry and Materials Science}, 
\orgname{Aalto University}, \orgaddress{\postcode{00076}, \city{Espoo}, \country{Finland}}}

\affil[4]{\orgdiv{Condensed Matter Physics Center (IFIMAC)}, 
\orgname{Universidad Aut\'onoma de Madrid}, \orgaddress{\postcode{28049}, \city{Madrid}, \country{Spain}}}

\affil[5]{\orgdiv{Nano Life Science Institute (WPI-NanoLSI)}, 
\orgname{Kanazawa University}, \orgaddress{\postcode{920-1192}, \city{Kanazawa}, \country{Japan}}}

\date{\today}

\begin{document}
\maketitle

\begin{abstract}

Predicting how organic molecules adsorb, assemble, and interact on metal surfaces is central to surface chemistry and molecular electronics, particularly in the context of interpreting high-resolution scanning probe microscopy. Yet, the application of first-principles simulations to interfaces is hampered by the computational cost for evaluating the electronic structure for the large number of atoms typically involved. 
We hereby present MAD-SURF, a machine learning interatomic potential specifically tailored for molecular adsorption on coinage metal surfaces. Trained on a broad dataset spanning diverse molecules, adsorption motifs, surfaces, molecular dynamics trajectories and non-covalent aggregates, MAD-SURF achieves accuracy comparable to the underlying DFT reference while enabling simulations orders of magnitude faster than density functional theory. The model reliably reproduces energies, forces and adsorption geometries across the three coinage metal substrates. 
We demonstrate its capabilities on experimentally characterized systems, including organic monolayers, polycyclic aggregates, flexible biomolecules and the long-range herringbone reconstruction of gold. By merging accuracy, speed, and generalizability, MAD-SURF offers a practical framework for accelerating atomistic simulations and advancing data-driven workflows in surface science.
\end{abstract}


\section*{Introduction}
Metal-organic interfaces are central to modern materials science~\cite{Liu2024Mar, Scott2003Mar}, providing the structural and electronic foundations for applications in catalysis~\cite{norskov2009towards}, molecular electronics~\cite{Heimel2007Apr, Vuillaume2003Nov, Dasari2011Jun}, and nanofabrication~\cite{Lugier2021Sep, Liu2017}. Density Functional Theory (DFT) is currently the main workhorse for modeling such interfaces, but its computational cost generally scales near cubically with system size, i.e.\ $\mathcal{O}(N^{3})$, due to the diagonalization of the Kohn-Sham Hamiltonian. In practice, this limits calculations from a few hundred up to around a thousand atoms even on modern supercomputers~\cite{keller2024small}. For more complex systems, such as large hydrocarbons in petroleum~\cite{Zhang2021Sep, chen_gross_acs_fuels}, molecular monolayers~\cite{Gardener2009Dec, Galeotti2020Aug, DeMarchi2019}, biomolecules~\cite{Grabarics2024Nov, Cai2025Mar, jestila_lignocellulosic, milica_stradiol}, or DNA~\cite{Pawlak2019Feb, Cai2022Nov}, the cubic scaling of DFT quickly renders it prohibitively costly, necessitating the development of faster and more scalable approaches while retaining its accuracy \cite{maurer_advances_2019}. Recently, machine learning interatomic potentials (MLIP) have emerged as a powerful alternative that combines near-DFT accuracy with orders-of-magnitude lower computational cost. They have successfully been applied across diverse domains of chemistry and materials science, from heterogeneous catalysis~\cite{oc2020, oc2022, adsorbml}, to biomolecular simulation~\cite{Noe2020Apr, microsoft_ai2bmd}, and even the discovery of new materials~\cite{mattersim, deep_mind_materials_discovery_nature}.

Despite these advances, most existing efforts have concentrated on crystalline solids~\cite{m3gnet_nat_comp_sci, mattersim, chgnet_nat_mach_int, deep_mind_materials_discovery_nature}, reactive adsorbates~\cite{oc2020, oc2022,stark_benchmarking_2024} or biomolecular systems in solution~\cite{Noe2020Apr, microsoft_ai2bmd, spice_nat_sci_data, ani_ml_jctc}. Notably, foundation models covering a wide array of materials including elements from most of the periodic table have recently emerged, providing pretrained potentials that can be finetuned to provide accurate properties for any system of interest with a much smaller dataset than when training from scratch~\cite{chgnet_nat_mach_int, batatia_foundation_2025, pet_mad_nat_comm}. However, these models are mainly trained on bulk materials, and analogous foundation models explicitly targeting dispersion-dominated, surface-screened adsorption and collective assembly of large organic adsorbates are currently lacking. Furthermore, polycyclic adsorbates stabilized primarily through non-covalent interactions at metal surfaces~\cite{Liu2014Nov} remain underrepresented in current MLIP benchmarks and training sets~\cite{westermayr_long-range_2022}, despite their importance in modern materials science. In this context, On-Surface Synthesis \cite{grill_nano-architectures_2007,clair_controlling_2019} combined with Scanning Probe Microscopy (SPM) has emerged as a unique tool to investigate these systems, enabling direct visualization of adsorption geometries, charge transfer, and even reactive intermediates at the atomic scale \cite{pavlicek2017,patera_probing_2025}. This capability has enabled detailed studies of catalytic reaction pathways~\cite{AlbrechtJACS2015, PavlicekNatChem2015}, single-molecule charge transport in molecular electronics~\cite{Bumm1996Mar, Fatayer2018May}, and biomolecular recognition at interfaces~\cite{Hinterdorfer2006May, Kienberger2006Jan}, providing insights not accessible to ensemble measurements. Nonetheless, despite the availability of efficient and accurate simulation tools~\cite{ellner2019molecular,  niko_oinonen_ppm_review}, interpretation for more complex metal-organic interfaces where van der Waals forces and conformational flexibility govern adsorption geometries and image contrast is still constrained by the limitations of DFT.
Accelerating the modeling of such systems while retaining near-DFT accuracy is essential for large-scale screening of potential adsorbate configurations~\cite{todorovic_bayesian_2019,egger_charge_2020,jestila_lignocellulosic}, in automating experimental workflows for molecular reactions~\cite{nian_jacs_spm} and discovery~\cite{Kalinin2021Aug, Li2025Jul}, as well as in terms of democratizing the computational sciences for researchers without access to extensive computational resources or high-performance computing facilities. 
Furthermore, recent work has shown that MLIPs can resolve long-standing problems in surface science out of scope for even DFT, in particular large-scale surface reconstruction phenomena. To this end, Li and Ding were able to explain the atomic-scale origin, thickness dependence and strain sensitivity of the Au(111) herringbone reconstruction with a DeePMD potential trained specifically for the Au surface, a system that is inaccessible to conventional \emph{ab initio} simulations owing to its large periodicity~\cite{li_origin_2022}. Accounting for such ubiquitous surface reconstructions in simulations provides inevitably more realistic models for catalysis~\cite{chen_machine-learning_2023}, self-assembly~\cite{sampath_closing_2020} and thin-film deposition~\cite{pieck_computational_2025}, and in combination with the extended time and length-scales at near-DFT accuracy, MLIPs represent a transformative tool for realistic large-scale MD simulations.    
Still, achieving near first-principles accuracy while simultaneously accessing the time and length scales required to describe adsorption dynamics, collective assembly, and thermal fluctuations at metal–organic interfaces remains a central challenge, particularly for large, flexible organic adsorbates stabilized by dispersion-dominated interactions on metal surfaces.

In this work, we introduce MAD-SURF (Molecular ADsorption on SURFaces), a machine learning interatomic potential tailored for molecular adsorption on coinage metal surfaces. MAD-SURF is trained on a diverse dataset specifically constructed for interfacial and on-surface chemistry, with all reference calculations performed at the Density Functional Theory level (PBE+vdW$^{\text{surf}}$). We focus on the widely used coinage metal surfaces (Cu, Ag, and Au), owing to their well-characterized electronic properties and distinct adsorption strengths. The dataset combines adsorption configurations obtained through active learning-based structure search, \emph{ab initio} molecular dynamics trajectories of molecules on these surfaces, normal mode sampling to capture intramolecular flexibility, and automated sampling of molecular clusters relevant to intermolecular interactions in aggregation and monolayer formation.
This multi-source strategy systematically covers local bonding, long-range dispersion, and conformational degrees of freedom, providing a robust foundation for training interatomic potentials for on surface chemistry.
Using this dataset, we benchmark different training strategies for the state-of-the-art MACE~\cite{batatia2022mace} architecture on their accuracy to predict energies, forces, adsorption heights, overall geometries, and transferability to larger adsorbates. From this comparison, the best-performing potential is selected, enabling simulations of extended organic adsorbates on surfaces with accuracy comparable to our DFT reference at a fraction of the computational cost. 
The resulting MAD-SURF potential provides a practical framework for simulating the adsorption, assembly, and dynamics of organic molecules on coinage metal surfaces with near first-principles accuracy. Beyond adsorption of isolated molecules, we demonstrate its generalization across chemically diverse systems, including organic monolayers, polycyclic molecular aggregates, and flexible biomolecules. Crucially, MAD-SURF enables long-time molecular dynamics simulations at system sizes and time scales that are inaccessible to DFT, allowing dynamic and collective phenomena such as thermal disorder, molecular rearrangement, and self-organization to be resolved directly at the atomistic level.
By bridging the gap between electronic-structure accuracy and experimentally relevant time and length scales, MAD-SURF facilitates quantitative interpretation of surface experiments and supports realistic simulations of complex metal–organic interfaces. We anticipate that the MAD-SURF model and dataset will empower both experimental and theoretical studies in surface chemistry, enable accelerated exploration of adsorption and assembly processes, and lower the barrier to high-fidelity atomistic modeling for systems beyond the reach of conventional \emph{ab initio} methods.

\section*{Results}

\begin{figure}[t!]
    \centering
    \includegraphics[width=0.9\linewidth]{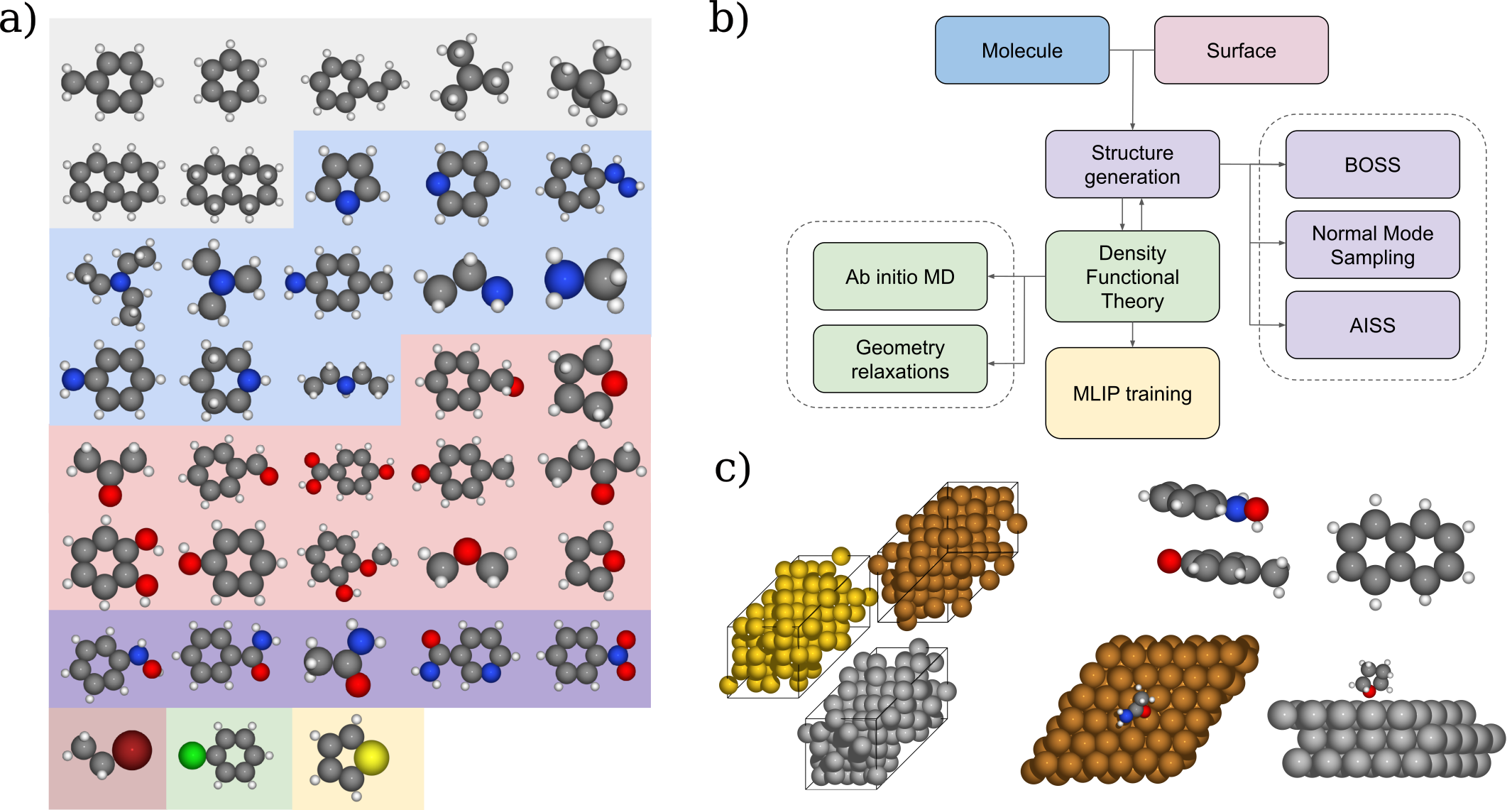}
    \caption{
    \textbf{Pipeline for the construction of the dataset.}
    (a) Representative molecules spanning functional groups relevant to on-surface chemistry, including aromatics, amines, carbonyls, halogens, hydroxyls, ethers, nitro groups, and extended $\pi$-conjugated systems. 
    (b) Computational workflow for dataset generation: molecular and surface structures are combined and sampled using Bayesian Optimization Structure Search (BOSS), Normal Mode Sampling (NMS), \emph{ab initio} Molecular Dynamics (AIMD), and Automated Interaction Site Screening (AISS). The resulting configurations are evaluated with Density Functional Theory to obtain reference energies and forces, and subsequently used for MLIP training. 
    (c) Example snapshots from the final dataset, illustrating MD trajectories of metals and molecules both in the gas phase and adsorbed.}
    \label{fig:overview_dataset}
\end{figure}

Training accurate and transferable machine learning interatomic potentials (MLIPs) requires a comprehensive dataset of reference (\emph{ab initio}) calculations. For on-surface chemistry, this dataset must capture not only equilibrium configurations but also out-of-equilibrium distortions, reactive intermediates, and diverse bonding motifs that occur across metal–organic interfaces. To achieve this, the MAD-SURF dataset spans a chemically broad and structurally diverse configuration space representative of realistic on-surface chemistry conditions. The MAD-SURF dataset includes molecules both in the gas phase and adsorbed on metal surfaces. To accurately reproduce the mechanical and thermodynamic response of adsorbates, it also incorporates intramolecular flexibility, intermolecular interactions, and distorted non-equilibrium structures, enabling the simulation of molecular aggregates and monolayers with partial and full coverage. Figure~\ref{fig:overview_dataset} summarizes the data generation pipeline. Panel (a) displays the 38 molecules included in our dataset, covering key functional groups relevant to on-surface chemistry: aromatics, amines, carbonyls, halogens, hydroxyls, ethers, nitro groups, and extended $\pi$-conjugated systems (see Table~S1). Panel (b) illustrates the computational workflow (see also Methods) combining pure DFT calculations such as \emph{ab initio} molecular dynamics and geometry relaxations with several complementary strategies for structure generation: Bayesian Optimization Structure Search (BOSS) for adsorption minima, Normal Mode Sampling (NMS) for intramolecular perturbations, and Automated Interaction Site Screening (AISS) for molecule–molecule interactions. Each configuration generated through the latter set of methods was subsequently re-evaluated with DFT to obtain accurate reference energies and forces.

The final composition of the dataset, summarized in Table~\ref{tab:dataset_hierarchy}, includes tens of thousands of configurations distributed across adsorption, molecular dynamics and gas-phase subcategories. Together, these data form a robust training basis for developing general MLIPs capable of describing a wide range of phenomena in on-surface chemistry, from single-adsorbate relaxation to collective self-assembly and high-temperature dynamics.

\begin{table}[b!]
\caption{Number of configurations per source method, system, and data modality, reported as \textbf{N\textsubscript{train} / N\textsubscript{test}}.}
\label{tab:dataset_hierarchy}
\centering
\begin{tabular}{lccc}
\toprule
Source method         & System / substrate & Data modality                 & \# configurations \\
\midrule
\rowcolor{bossbg}
BOSS                    & Cu(111)    & Bayesian optimization & 3348 / 372 \\
\rowcolor{bossbg}
BOSS                    & Cu(111)    & Relaxation            & 5504 / 612 \\
\rowcolor{bossbg}
BOSS                    & Ag(111)    & Relaxation            & 1299 / 146 \\
\rowcolor{bossbg}
BOSS                    & Au(111)    & Relaxation            & 4713 / 524 \\
\midrule
\rowcolor{mdsurfbg}
MD molecule/substrate   & Cu(111)    & Molecular dynamics    & 4602 / 512 \\
\rowcolor{mdsurfbg}
MD molecule/substrate   & Ag(111)    & Molecular dynamics    & 15968 / 1774 \\
\rowcolor{mdsurfbg}
MD molecule/substrate   & Au(111)    & Molecular dynamics    & 11405 / 1268 \\
\midrule
\rowcolor{mdmetalbg}
MD metals               & Cu         & Molecular dynamics    & 4056 / 451 \\
\rowcolor{mdmetalbg}
MD metals               & Ag         & Molecular dynamics    & 6833 / 760 \\
\rowcolor{mdmetalbg}
MD metals               & Au         & Molecular dynamics    & 4682 / 521 \\
\midrule
\rowcolor{aissbg}
AISS                    & Gas-phase  & Static                & 28752 / 3195 \\
\midrule
\rowcolor{nmsbg}
NMS                     & Gas-phase  & Static                & 1278 / 144 \\
\bottomrule
\end{tabular}
\end{table}

\subsection*{Benchmarking training strategies}
The training of machine learning interatomic potentials, especially for models that are expected to generalize beyond their training domain, is currently an active area of research and therefore we systematically evaluated multiple strategies, as described in more detail in the Methods. Broadly speaking, we can divide these strategies into $\lambda$-tests (with $\lambda$ being the force/energy weight ratio in the loss function used for training), naïve or descriptor-based filtering (filtering either by taking every N-th data entry in the set or by estimating their similarity using the MACE-descriptor), and finetuning from the foundational model.

\begin{figure}[t!]
    \centering
    \includegraphics[width=1\linewidth]{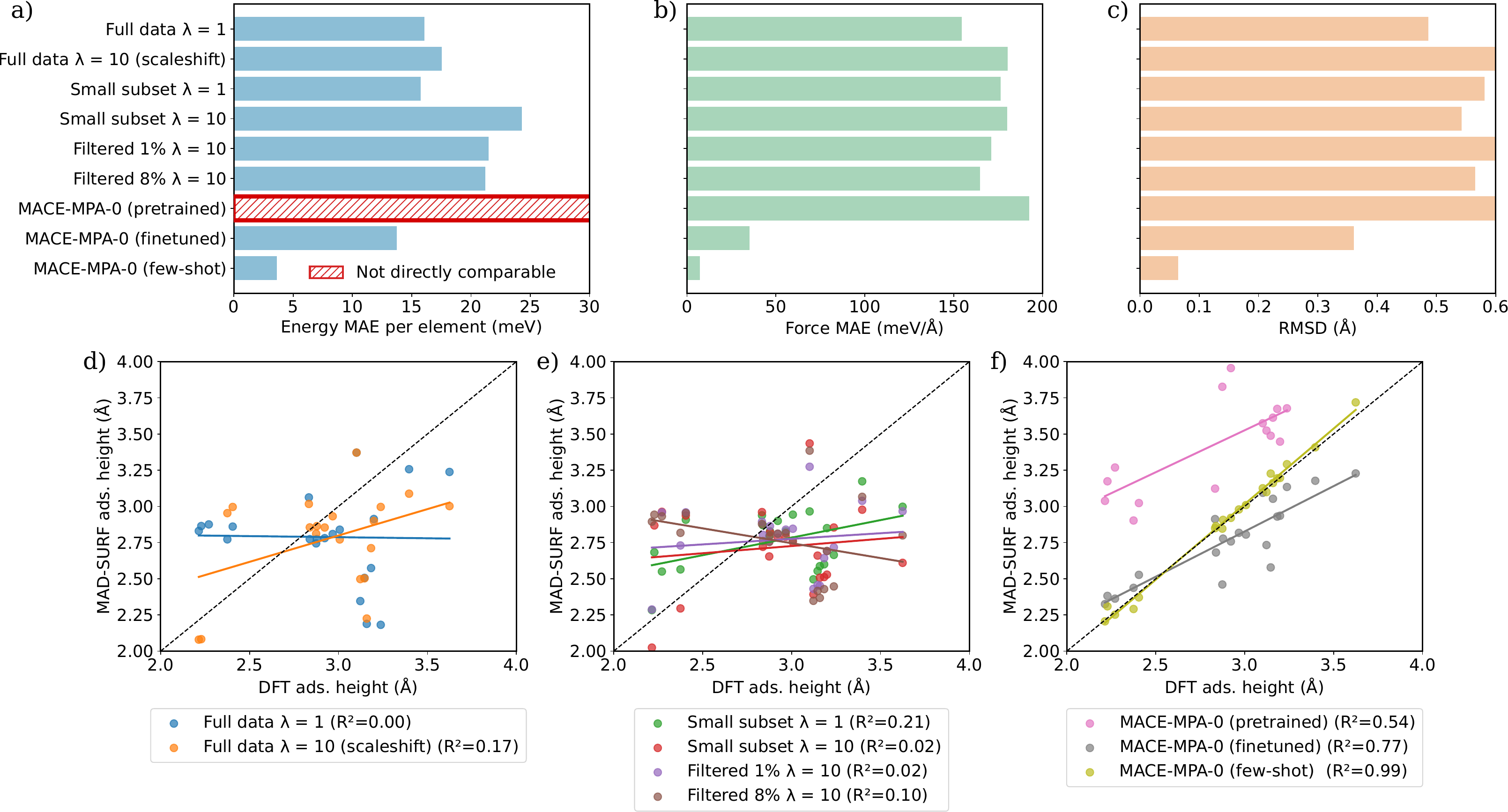}
    \caption{
    \textbf{Comparison of different training strategies for a subset containing the best of each type.} 
     (a) Plot of the metrics used to evaluate the trained potentials: mean absolute error (MAE) in per-atom energies (meV), forces (meV/Å), adsorption height differences and RMSD values (Å) between DFT reference and MAD-SURF geometries, sorted by increasing RMSD values. (b) Parity plot over the adsorption heights (Å) of the test molecules on Cu(111), Ag(111) and Au(111) for the DFT reference and MAD-SURF geometries. Note that the per-atom energy MAE for the MACE-MPA-0 foundational model is based on plane-wave DFT (VASP), which cannot be compared directly against our reference based on localized basis sets (FHI-aims). 
    }
    \label{fig:mlip_metrics}
\end{figure}

For the tests themselves, we wanted to ensure the reliability of our trained potentials during simulations, and thus we evaluated them based on how closely the MLIP brings the systems, in terms of structure, energies and forces, to the reference DFT method during relaxation. The results are displayed in Fig.~\ref{fig:mlip_metrics}a-c, where the per-atom energy mean absolute errors (MAE) and force MAEs of MAD-SURF are reported relative to the DFT-relaxed structures, whereas the element-weighted root-mean-squared deviations (RMSD) are reported between the MAD-SURF and DFT structures. The parity plots in Fig.~\ref{fig:mlip_metrics}d-f show the discrepancy between the adsorption-height trends as determined by MAD-SURF and DFT. The training was initialized on the \textit{full} dataset without any filtering, followed by training on the \textit{small} subset. Here we found that in terms of the energy MAEs both models trained using $\lambda$=1 display similar errors. Meanwhile, there is a slightly larger discrepancy between the \textit{full} and \textit{small} $\lambda$=10 models. Furthermore, the force MAEs and the RMSDs are highly similar for these, providing little indication as to which models are more reliable. In contrast, when combining these metrics with the adsorption height parity, we note that even though the energy MAEs suggest slightly improved model performance for the \textit{full} model, none of the aforementioned models are successfully capturing the correct adsorption heights trends. The situation was not ameliorated by the more refined filtering methods, leading to similar low correlation. 

Motivated by the limited adsorption-height fidelity of the models trained from scratch, we next evaluated transfer learning starting from the pretrained MACE-MPA-0 foundational model~\cite{batatia_foundation_2025}. As a reference point, the unmodified pretrained model (MACE-MPA-0 (pretrained)) yields only moderate agreement for adsorption heights ($R^{2}=0.54$) and exhibits a systematic tendency to overestimate adsorption distances.
This behavior is consistent with its pretraining being dominated by bulk materials data, rather than by dispersion-dominated, surface-screened adsorption of large organic molecules. We therefore performed finetuning on our MAD-SURF dataset (MACE-MPA-0 (finetuned)). This finetuning substantially improves adsorption-height trends and structural fidelity, increasing the adsorption-height correlation to $R^{2}=0.77$ and reducing errors across the evaluated metrics (Fig.~\ref{fig:mlip_metrics}). 
We selected this model for all subsequent simulations in this work.
In addition, we explored a practical \emph{few-shot} finetuning intended for rapid, molecule-specific refinement in experimental workflows. 
Here, we perform an additional lightweight adaptation of the foundational model using a small set of static reference calculations for a couple of target adsorbates, subsequently evaluating the calibrated model on additional configurations of the same adsorbates not included in the adaptation set.
As expected, this molecule-specific calibration yields highly accurate adsorption heights for the calibrated adsorbates, reaching $R^{2}=0.99$ in our benchmark (MACE-MPA-0 (few-shot)). 
We emphasize that this few-shot calibration is an application mode rather than a measure of general transferability and it is reported as an upper bound on accuracy. 

\subsection*{Molecular aggregates}

\begin{figure}[t!]
    \centering
    \includegraphics[width=0.8\linewidth]{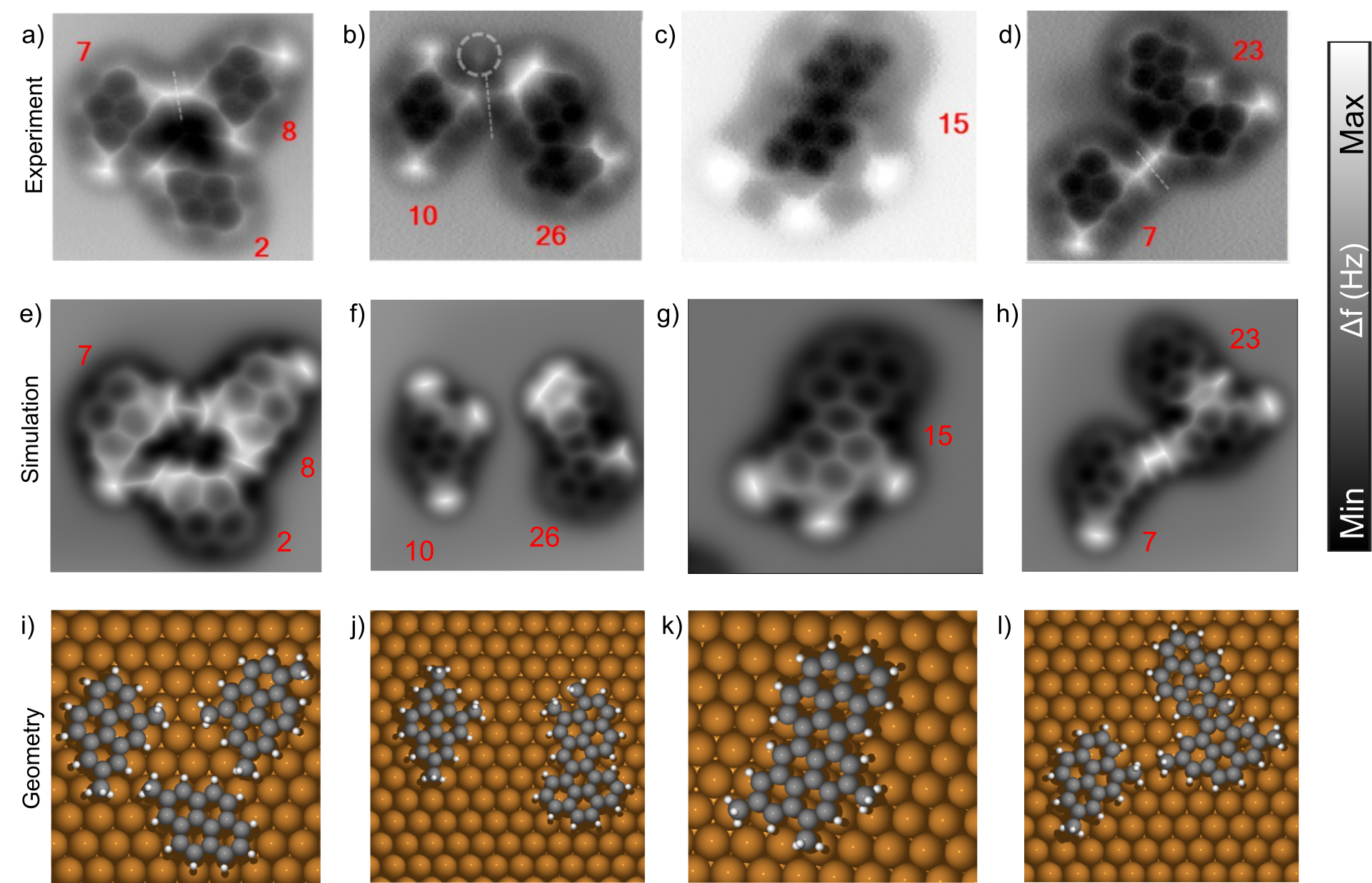}
    \caption{
    \textbf{Aggregated aromatic hydrocarbons from petroleum on Cu(111).} 
    NC-AFM experimental (a–d) images and simulations with the Probe Particle Model~\cite{HapalaPRB2014_ppm} (Lennard–Jones plus point charges) (e–h) of different thermalized products from 2,7-dimethylpyrene adsorbed on Cu(111). 
    (i–l) Top view  of the atomic structures used for the AFM simulations, obtained through our MLIP-based relaxation pipeline. 
    Experimental images adapted from Chen et al.~\cite{chen_gross_acs_fuels}. 
}

    \label{fig:leo_gross}
\end{figure}

To evaluate the applicability of MAD-SURF beyond isolated adsorbates, we next examine its ability to reproduce experimentally observed molecular aggregates. This test directly addresses whether the potential can recover adsorption geometries consistent with experimental high-resolution SPM images, thereby bridging imaging and atomistic interpretation at minimal computational cost. We take as a benchmark the study by Chen et al.~\cite{chen_gross_acs_fuels}, where several thermally induced reaction products of 2,7-dimethylpyrene were imaged on Cu(111) using Atomic Force Microscopy (AFM). In that work, the molecular structures were inferred from experimental contrast patterns, but the detailed adsorption geometries and molecule–substrate registries were not computed. Using MAD-SURF, we reconstructed the atomic arrangements corresponding to the experimental images shown in Fig.~\ref{fig:leo_gross}. We first optimized the gas-phase molecules with MAD-SURF and then placed them on the Cu(111) surface in initial orientations consistent with the experimental motifs. Each system was relaxed following the same convergence criteria as for the DFT geometries (see Methods). Subsequently, the relaxed structures were used to generate constant-height AFM simulations with the Probe Particle Model~\cite{HapalaPRB2014_ppm, niko_oinonen_ppm_review}.

The resulting simulated images closely match the experimental contrast, reproducing both intra- and intermolecular features, validating the model’s ability to reproduce adsorption geometries and molecule–molecule interactions. Overall, this case study demonstrates that MAD-SURF can efficiently recover realistic adsorption configurations for complex thermally generated polycyclic aggregates, achieving image-level agreement with experiment at a fraction of the computational cost of conventional DFT. This ability to interpret AFM data highlights MAD-SURF’s potential as a practical tool for accelerating experimental workflows for the SPM community.

\subsection*{Organic monolayers}

To validate the performance of MAD-SURF on extended organic adsorbates, we first examined its ability to reproduce experimentally observed monolayer structures, which are largely governed by a delicate balance between intermolecular and surface interactions. We study the prototypical systems of the herringbone pentacene/Cu(111) monolayer~\cite{smerdon_monolayer_2011} and various PTCDA monolayer phases on Cu(111), Ag(111) and Ag(110) surfaces~\cite{wagner_initial_2007, glockler_highly_1998}, which have been extensively characterized using SPM.

\begin{figure}[t!]
    \centering
    \includegraphics[width=0.7\linewidth]{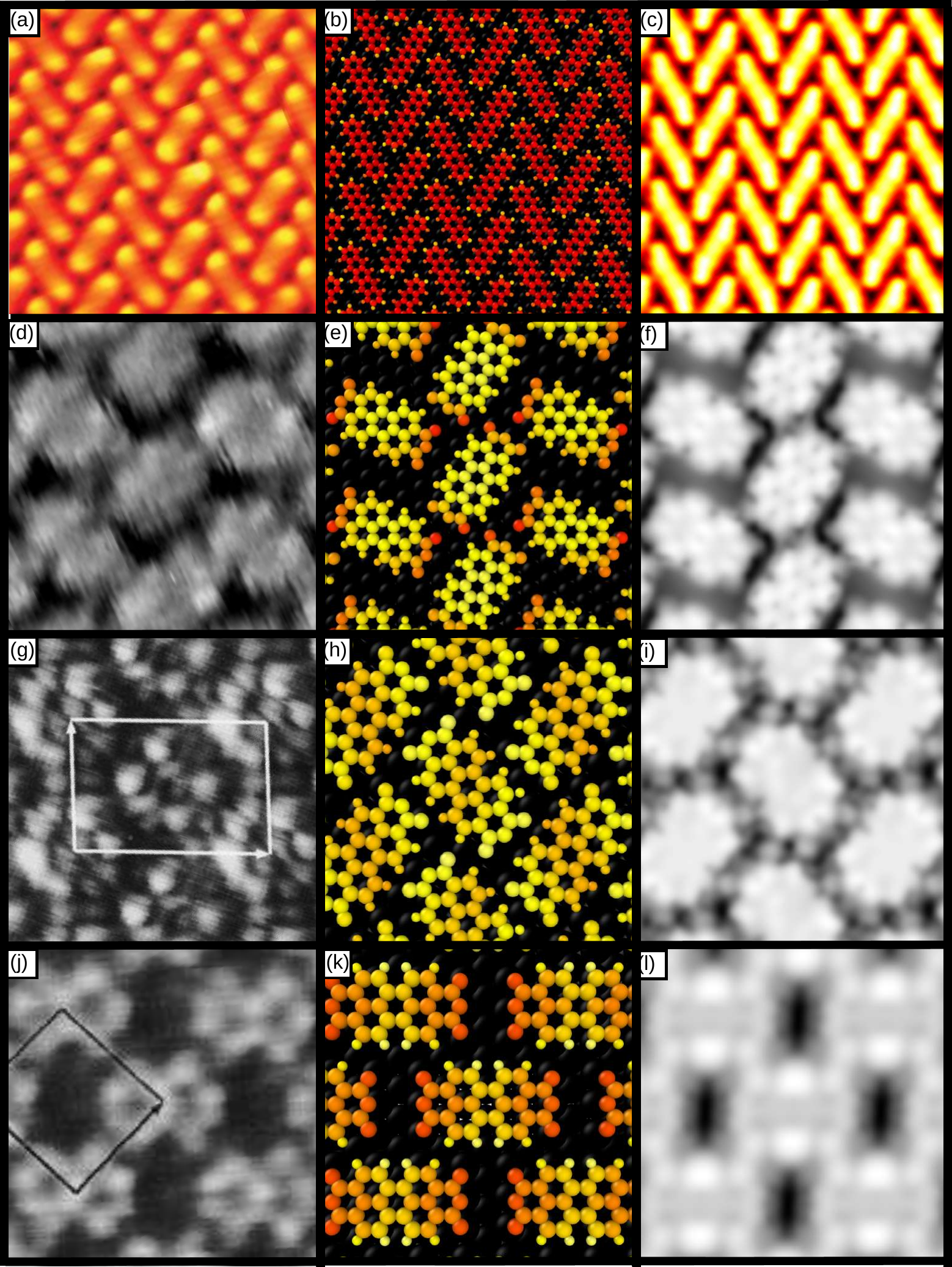}
\caption{
\textbf{Organic monolayer domains of pentacene and PTCDA on coinage metal surfaces.}
Rows from top to bottom correspond to the herringbone monolayer of pentacene on Cu(111) (a-c), the herringbone phase of PTCDA on Cu(111) (d-f), the herringbone phase of PTCDA on Ag(111) (g-i), and the brick-wall phase of PTCDA on Ag(110) (j-l). For each system, the left panel shows an experimental Scanning Tunneling Microscopy (STM) topograph, the middle panel the MAD-SURF–relaxed structural model with atoms color-coded by height, and the right panel a simulated STM image obtained using the Tersoff–Hamann approximation~\cite{TersoffHamann} and the same tunnelling bias as in the corresponding experiment. Experimental STM images were adapted from Smerdon et al.~\cite{smerdon_monolayer_2011} for pentacene/Cu(111) (15×15 nm$^{2}$, $V_\mathrm{gap}=0.5$ V, $I_\mathrm{T}=0.6$ nA), from Wagner et al.~\cite{wagner_initial_2007} for PTCDA/Cu(111) ($V_\mathrm{gap}=0.6$ V, $I_\mathrm{T}=10$ pA), and from Glöckler et al.~\cite{glockler_highly_1998} for PTCDA/Ag(111) ($V_\mathrm{gap}=2.0$ V, $I_\mathrm{T}=0.6$ nA) and PTCDA/Ag(110) ($V_\mathrm{gap}=0.9$ V, $I_\mathrm{T}=1.3$ nA).
}
    \label{fig:monolayers_stm}
\end{figure}

As shown in Fig.~\ref{fig:monolayers_stm}, MAD-SURF yields monolayer structures that are generally in good qualitative agreement with experiment. The potential captures the fine balance between intermolecular hydrogen bonding interactions, adsorbate-substrate interactions, and intramolecular relaxation of the adsorbates resulting from these. The most notable differences between the MAD-SURF structures and the DFT reference can be seen in the pentacene on Cu(111) and the herringbone phase of PTCDA on Ag(111). In the case of the former, the pentacene molecules are less distorted by the substrate with MAD-SURF, and are slightly more planar (0.2 vs. 0.4 Å between lowest and highest atoms) than with the DFT reference (Fig.~S10). Meanwhile, for the PTCDA herringbone phase on Ag(111), the anhydride groups at the edges of the molecules are not sufficiently distorted towards the surface (Fig.~S9). 
Furthermore, both dispersion-corrected DFT and MAD-SURF predict a higher adsorption height for PTCDA on Ag(111) than observed experimentally ($\sim$3.00~\AA{} vs.\ 2.86~\AA{}). This discrepancy is consistent with known limitations of semi-local DFT combined with pairwise dispersion corrections for this system, and is therefore inherited by the MLIP through its training on the DFT reference. At the theoretically predicted adsorption distance, the PTCDA LUMO lies at the Fermi level, whereas constraining the molecule to the experimental height shifts the LUMO by 0.2--0.3~eV below the Fermi level, leading to a qualitatively different STM contrast~\cite{emiliano_brstm}. Similar sensitivities of the potential energy surface and electronic structure to the adsorption height have been reported for PTCDA/Ag(111) across different vdW approaches~\cite{Hormann2020}. Other than these differences, MAD-SURF provides structures that reproduce the main features of experimentally characterized monolayers, with adsorption structures and heights close to experimental and DFT results. A quantitative comparison of adsorption heights for pentacene and PTCDA monolayers on Cu and Ag surfaces is reported in Table~\ref{tab:monolayer_heights}, showing deviations below 0.2 Å in all cases. 

\begin{table}[h!]
\caption{Adsorption heights of pentacene and PTCDA monolayers on coinage metal surfaces obtained with MAD-SURF compared to experimental reference values~\cite{koch_adsorption-induced_2008,gerlach_substrate-dependent_2007, mercurio_adsorption_2013}.}

\label{tab:monolayer_heights}
\centering
\begin{tabular}{lcccc}
\toprule
System & Surface & MAD-SURF (Å) & Experiment (Å) & $|\Delta z|$ (Å) \\
\midrule
Pentacene & Cu(111) & 2.16 & $2.34 \pm 0.02$ & 0.18 \\
PTCDA & Cu(111) & 2.78 & $2.66 \pm 0.02$ & 0.12 \\
PTCDA & Ag(111) & 3.01 & $2.86 \pm 0.01$ & 0.15 \\
PTCDA & Ag(110) & 2.44 & $2.52 \pm 0.11$ & 0.08 \\
\bottomrule
\end{tabular}
\end{table}

\subsection*{Biomolecules}

\begin{figure}
    \centering
    \includegraphics[width=0.8\linewidth]{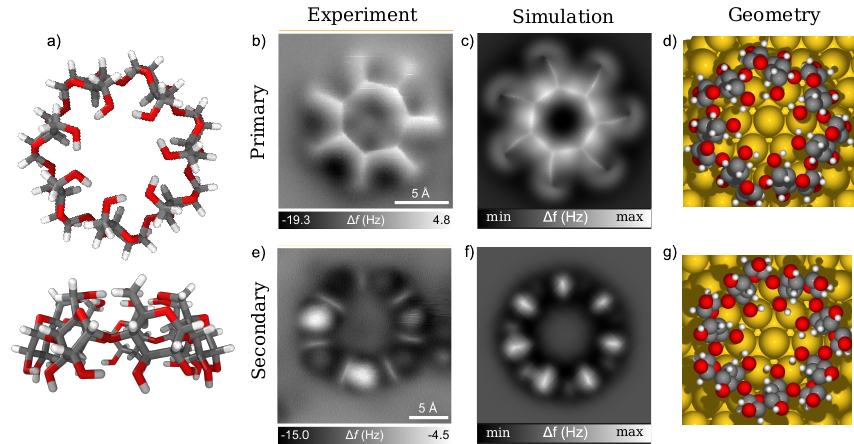}
    \caption{
    \textbf{Biomolecular adsorbate $\beta$-cyclodextrin on Au(111).} 
    (a) Top and side views of the gas-phase structure, highlighting the two faces that can adsorb on the substrate. 
    (b,c) Experimental and simulated constant-height images of $\beta$-cyclodextrin adsorbed on Au(111) exposing the primary-face orientation, where the narrow rim forms a closed hydrogen-bonding network. 
    (d,e) Experimental and simulated images of the secondary-face orientation, defined by the seven secondary OH groups forming the wider rim. 
    (f,g) Top and side views of the atomic structures used for the AFM simulations, obtained through our MLIP-based relaxation pipeline.  AFM simulations performed with the  Probe Particle Model~\cite{HapalaPRB2014_ppm} (Lennard–Jones plus point charges)
    Experimental images adapted from Grabarics et al.~\cite{Grabarics2024Nov}. 
}

    \label{fig:betacd}
\end{figure}

Biomolecules play vital roles in biological recognition, catalysis, and materials design, but their intrinsic structural complexity and conformational flexibility make it difficult to resolve their atomic-scale structures using conventional spectroscopic or diffraction techniques. In this context, SPM can overcome these limitations, enabling direct visualization of individual functional groups and hydrogen-bonding networks within single molecules. A representative case is $\beta$-cyclodextrin ($\beta$-CD), a cyclic heptamer of glucose units forming a truncated cone with two distinct rims: the narrow primary face and the wider secondary face. These two orientations can be discriminated experimentally using high-resolution AFM, resolving individual hydroxyl groups and the associated hydrogen-bond network~\cite{Grabarics2024Nov}. 
Building on this, we employed MAD-SURF to reproduce the adsorption geometries of $\beta$-CD on Au(111) and their corresponding AFM contrasts (Fig.~\ref{fig:betacd}).
Instead of the extensive molecular dynamics sampling used in the original work of Grabarics et al.~\cite{Grabarics2024Nov}, we adopted a simpler workflow: conformational sampling was performed with CREST~\cite{pracht_automated_2020} at the tight-binding level to identify the most stable gas-phase conformer, which was then adsorbed on both the primary and secondary faces using MAD-SURF. Subsequently, a 10.0 ps annealing of the structures was performed, decreasing the temperature from 420 to 153 K during the simulation, which allowed the cyclic hydrogen bonding networks to fully align in either clockwise or counterclockwise directions. To avoid drifting of the molecules on the surface during annealing, we fixed all atoms up to the lowest carbon atoms in the rims. The relaxed structures were used to compute constant-height AFM images.
For the primary-face adsorption geometry, the simulated constant-height AFM image is in qualitative agreement with experiment (Fig.~\ref{fig:betacd}c,d), resolving the closed heptagonal motif of the intramolecular OH hydrogen-bond network, with the C–O bonds appearing as radial arms. In contrast, for the secondary-face orientation the simulation produces seven lobes of comparable repulsive contrast, whereas the experiment exhibits two markedly brighter lobes.
We attribute this discrepancy to the experimental environment: the imaged molecule is embedded in a densely packed island rather than isolated, where the lateral intermolecular interactions and/or local packing constraints breaks the sevenfold symmetry.
Finally, we tested whether the clockwise versus counterclockwise orientation of the rim OH network yields a measurable contrast difference (SI: $\beta$-CD OH orientation analysis). Both Probe Particle Model simulations (Lennard–Jones plus point charges) and Full Density-Based Model calculations~\cite{ellner2019molecular} (Fig.~S7) predict indistinguishable contrast for the two orientations.

\subsection*{Au(111) herringbone reconstruction}

The Au(111) herringbone reconstruction is a prototypical stress–relief pattern of the clean gold surface, characterized by a $\sim$30 nm periodic network of alternating rotational domains with fcc and hcp stacking domains separated by dislocation lines, in turn separated by elbows. Although it is routinely observed in STM experiments, it has long been difficult to reproduce in atomistic simulations. Conventional DFT slab calculations are limited to comparatively thin slabs and small lateral cells, which prevents them from capturing the long-range elastic relaxation and subtle energetics that stabilize the reconstruction over the unreconstructed $(1\times 1)$ or $22\times\sqrt{3}$ stripe phases. Classical or semiempirical force fields, in turn, struggle to describe the delicate balance between surface stress, subsurface strain and stacking faults that gives rise to the pattern. Recently, Li and Ding resolved this problem by constructing a dedicated, gold-specific machine-learning force field through an active-learning workflow based on the DeePMD~\cite{li_origin_2022} architecture. In their approach, MLIP–MD simulations, DFT reference calculations and model retraining were iterated to systematically target bulk elastic properties, vacancy and adatom energetics and candidate reconstructed Au(111) slabs. After this focused data acquisition and optimization, yielding a highly specialized single-element potential, where large-scale simulations with more than 30 atomic layers and tens of thousands of atoms recover the correct periodicity and strain response of the herringbone phase.

\begin{figure}
    \centering
    \includegraphics[width=0.8\linewidth]{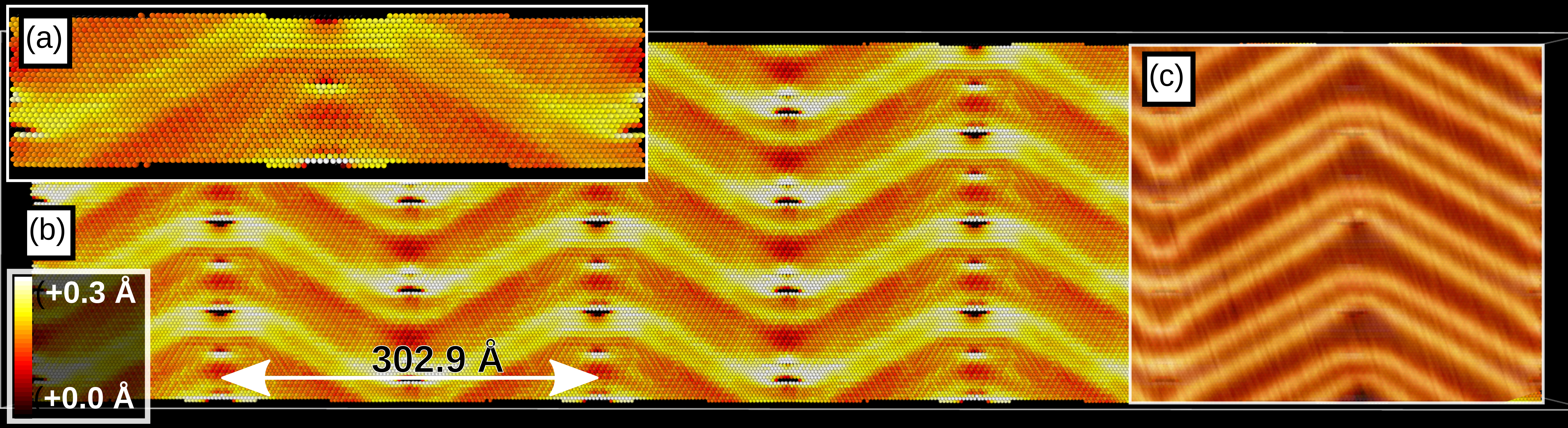}
    \caption{
    \textbf{Au(111) herringbone surface reconstruction.} 
    (a) Unit cell (302.9$\times$71.7 Å$^2$) of the Au(111) herringbone reconstruction with relaxed lattice vectors and atomic positions acquired using MAD-SURF. 
    (b) 4$\times$4 supercell with the emerging herringbone pattern, color coded by surface layer corrugation in Å. 
    (c) Experimental STM image, adapted from Walen et al.~\cite{walen_self-organization_2015}. 
}
    \label{fig:au111_hb}
\end{figure}

Here we use the Au(111) herringbone reconstruction as a stringent test for our general-purpose MLIP trained for on-surface chemistry to examine whether it can reproduce such a subtle metal-only reconstruction without any additional Au-specific active learning. Starting from the same slightly densified Au(111) stripe pattern unit cell as reported in~\cite{hanke_structure_2013}, we generated and aligned the two herringbone rotational domains, in similar fashion as Li and Ding~\cite{li_origin_2022}. We then used MAD-SURF in exactly the same form as for the molecule–surface calculations, and the system relaxes into a herringbone-like geometry whose periodicity and elbow morphology resemble experimental STM images (Fig.~\ref{fig:au111_hb}). A commonly  reported value for the periodicity\textemdash corresponding to the distance between elbows along the horizontal direction in the Figure (approximately parallel to [01$\overline{1}$])\textemdash is 324.3 Å at experimental temperatures between 300 and 700 K~\cite{sandy_structure_1991, voigtlander_epitaxial_1991}, while our value is slightly shorter at 302.9 Å. The experimental surface corrugation is around 0.2 Å~\cite{li_origin_2022}, which is in fair agreement with our MAD-SURF model (see color scale in Fig. \ref{fig:au111_hb}). Note however that our model has an additional edge dislocation per repeating unit in each rotational domain, while there should be only one. Due to this slight misalignment of rotational domains, there is an additional corrugated feature in the fcc domain between elbows. This proof-of-concept indicates that the Au component of MAD-SURF captures essential aspects of the long-range elastic response involved in the reconstruction, despite not being specifically optimized for this system.

\subsection*{Large-scale dynamics at experimentally relevant scales}

\begin{figure}[t!]
    \centering
    \includegraphics[width=0.8\linewidth]{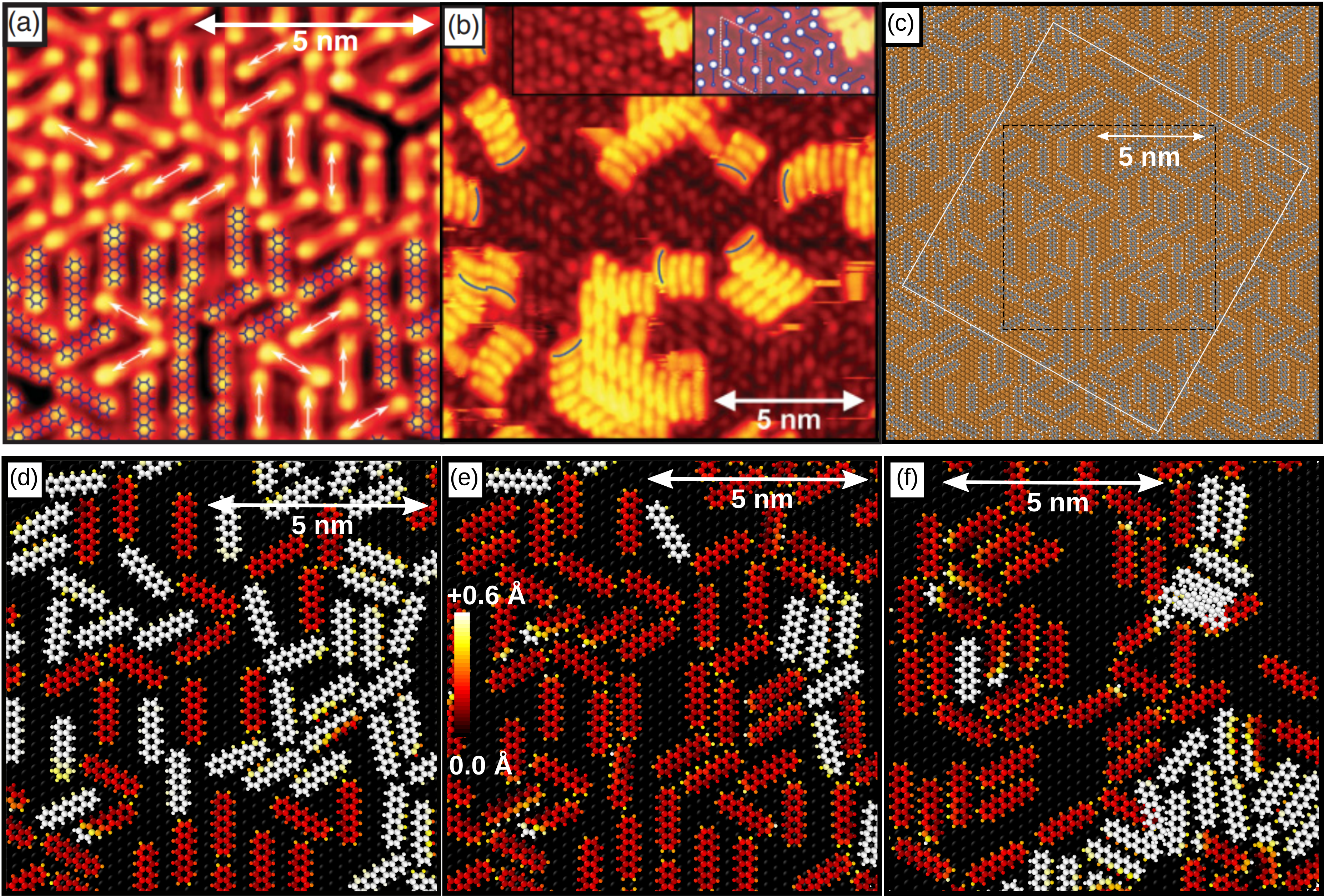}
    \caption{
    \textbf{Example of large-scale dynamics of organic monolayers: the random-tiling (R) monolayer structure of pentacene on Cu(111).} 
    (a) 10$\times$10 nm$^{2}$ STM topograph (\textit{V}$_{gap}$ = 0.5 V, \textit{I$_{T}$} = 1 nA) with superimposed molecules on a portion of the monolayer, (b) 15$\times$15 nm$^{2}$ STM topograph (\textit{V}$_{gap}$ = 2.0 V, \textit{I$_{T}$} = 0.6 nA) of pentacene bilayer structures where the molecules are tilted about their long axes, (c) the initial guess for the large-scale 13.27$\times$13.71 nm$^{2}$ model structure (118 Pn molecules, 17144 total atoms) used in our MLIP simulations, the black dashed box corresponding to the 10$\times$10 nm$^{2}$ portion shown in (a-b) and (d-f), while the solid white box corresponds to the unit cell of the periodic model. Pentacene random-tiling model structures with atoms color-coded according to height (z-coordinate) to indicate and emulate the image contrast in the corresponding STM topographs, obtained by: (d) relaxing the initial guess structure, (e) relaxation following a 100 ps MD run at 55 K, and (f) relaxation following a 50 ps MD run at 420 K and a subsequent 50 ps cooldown to 55 K. Experimental images and structure models (a-b) adapted from Smerdon et al.~\cite{smerdon_monolayer_2011}. 
    }
    \label{fig:pn_R_phase}
\end{figure}

Currently there are few tools capable of dynamic simulations with accuracy comparable to electronic-structure methods for systems containing more than a few molecules on an extended surface area. Furthermore, static and dynamic DFT computations are often constrained to highly symmetric repeating patterns, single-crystal bulk structures and well-defined low Miller index surfaces~\cite{montoya_challenge_2015}, limiting their utility for simulating stochastic processes dominated by random structures and disorder at experimentally relevant scales. Simultaneously, classical MD is up to the task in terms of scale, yet cannot be used to model reactions and complex environments accurately. MLIPs such as MAD-SURF merge the best of DFT and classical MD, providing favorable scaling and quantum accurate properties, and thus holds promise to revolutionize how and what can be simulated within heterogeneous catalysis~\cite{chen_machine-learning_2023}, self-assembly~\cite{sampath_closing_2020}, thin-film deposition~\cite{pieck_computational_2025}, and functional materials such as phase-change systems and electrochemical energy-storage electrodes~\cite{deringer_machine_2019}. Here, we demonstrate the application of MAD-SURF to the simulated annealing of a random phase of 118 pentacene molecules, requiring a 13.27$\times$13.71 nm$^{2}$ unit cell (Fig. \ref{fig:pn_R_phase}c) to capture the 10$\times$10 nm$^2$ area in the experimental image (Fig. \ref{fig:pn_R_phase}a). Following a brief MD simulation at the STM imaging temperature (55 K), where the molecules are mostly immobile, we only observe minor oscillations of the molecules around their equilibrium adsorption positions. Furthermore, the adsorption heights (MAD-SURF: 2.30 $\pm$ 0.10 \AA, Exp: 2.34 $\pm$ 0.02 \AA)~\cite{koch_adsorption-induced_2008}, slightly concave geometries and relative molecular positions are also consistent with experiment (Fig. \ref{fig:pn_R_phase}e). Following an increase in the simulation temperature to 420 K, a new random phase is established (Fig. \ref{fig:pn_R_phase}f), accompanied by emerging bilayers consistent with experiment (Fig. \ref{fig:pn_R_phase}b). These results demonstrate how MAD-SURF enables dynamic simulations of phenomena previously unattainable with the same combined scale and accuracy, and can act as solid foundation for more specialized reactive MLIPs through finetuning.

\section*{Conclusions}

In this work, we have introduced MAD-SURF, a general-purpose machine learning interatomic potential (MLIP) for molecular adsorption on coinage metal substrates relevant to on-surface chemistry. MAD-SURF is trained on a large and chemically diverse dataset calculated at the Density Functional Theory reference level (PBE+vdW$^{\text{surf}}$), specifically parametrized for surface adsorption. The multi-source strategy used for building the dataset spans local bonding, long-range dispersion and intramolecular flexibility. On this basis, MAD-SURF achieves accuracy comparable to our PBE+vdW$^{\text{surf}}$ reference for energies, forces and adsorption heights, while reducing the computational cost by several orders of magnitude. We systematically benchmarked different training strategies for the MACE architecture, including variations in the energy–force loss balance, naïve and descriptor-based filtering of configurations, and finetuning of a foundational model. These tests show that conventional error metrics on energies and forces alone are not sufficient to guarantee reliable adsorption geometries, and that these, in combination with adsorption-height parity, offer a more stringent and physically meaningful indicator of MLIP quality for molecular adsorption on surfaces. In particular, we found that finetuning the MACE-MPA-0 foundational model substantially improves adsorption-height correlations and structural fidelity, while retaining good force accuracy and MD stability. From this finetuned model, we demonstrated transferability across several classes of experimentally relevant systems. MAD-SURF reproduces the adsorption geometries of organic monolayers such as pentacene and PTCDA on Cu(111), Ag(111) and Ag(110), captures the adsorption structures of thermally generated polycyclic molecular aggregates from petroleum, and accurately describes the adsorption and hydrogen-bonding patterns of the flexible biomolecule $\beta$-cyclodextrin on Au(111). Beyond molecule–surface systems, the same MLIP recovers the Au(111) herringbone reconstruction without any Au-specific active learning, indicating that the metal components of MAD-SURF faithfully encodes the long-range elastic response required for complex surface reconstructions. Finally, we showed that MAD-SURF enables large-scale finite-temperature simulations of disordered organic monolayers, exemplified by the random-tiling phase of pentacene on Cu(111) in experimentally sized cells. These simulations, which are out of reach for conventional DFT-based dynamics, capture adsorption heights, molecular conformations and the emergence of bilayers in good agreement with experimental observations. 
Together, these results demonstrate that MAD-SURF merges near DFT-level accuracy while reaching time and length scales that are typically only accessible to classical MD, thereby enabling dynamic simulations of complex on-surface phenomena with high fidelity. By providing both the MAD-SURF dataset and pretrained potential, we aim to lower the barrier for development of new MLIP tailored for on-surface chemistry and to accelerate structure determination and dynamics simulation in surface science, particularly in workflows that couple SPM experiments with atomistic modeling. In future work, the framework established here can be extended by enlarging the chemical and materials space to additional substrates, defects and adsorbates, and by finetuning towards reactive events and chemisorption. We anticipate that MAD-SURF and related general MLIPs will become key components in data-driven and autonomous approaches to on-surface synthesis, heterogeneous catalysis and molecular nanofabrication.

\clearpage

\section*{Methods} \label{methods}

\subsection*{Density functional theory (DFT)}
All reference calculations were performed with the all-electron numeric atom-centered orbital code FHI-aims~\cite{blum_ab_2009, HavuV09, RenX12, Levchenko15}. 
We used the PBE exchange–correlation functional~\cite{PBE1996} augmented by the Tkatchenko–Scheffler dispersion correction~\cite{tkatchenko_accurate_2009} with the surface-screened parameterization (DFT+vdW$^{\text{surf}}$)~\cite{ruiz_density-functional_2012, ruiz_density-functional_2016}. 
Scalar-relativistic effects were treated within the atomic ZORA approximation~\cite{van_lenthe_relativistic_1994, van_lenthe_zero-order_1996}, and all calculations employed the ``tight'' basis defaults unless stated otherwise. 
Geometry optimizations were carried out until the maximum force on any atom was below 0.01 eV\,Å$^{-1}$.
For AIMD simulations, we used the canonical ensemble (NVT) with a Berendsen thermostat~\cite{berendsen1984molecular}, a time step of 0.5 fs, and a total simulation length of 5 ps. 

\subsection*{Dataset generation}
We constructed a diverse dataset covering different small- to medium-sized molecules, metal surfaces, and their combinations, both in and out of equilibrium. On these surfaces we placed a curated set of small to medium-sized molecules that encompass the most common functional groups found in SPM-relevant organic adsorbates~\cite{GonzalezLastre2024Dec,QUAM-AFM_JCIM}. The aim was to systematically cover the chemical motifs that dominate molecule–surface interactions: amines, carbonyls, aromatic nitrogen heterocycles, halogens, hydroxyls, ethers, nitro groups, and extended aromatic systems (Fig.~S1-S2), encompassing the elements H, C, N, O, S, Cl, Br, Cu, Ag, Au. This design ensures that our training data span hydrogen bonding, $\pi$–$\pi$ stacking, metal–heteroatom coordination, and van der Waals physisorption. The dataset integrates four complementary sources:  
(i) adsorption structures from Bayesian Optimization Structure Search (BOSS),  
(ii) AIMD trajectories of adsorbates on coinage-metal surfaces and of bulk/surface slabs with and without vacancies,  
(iii) intramolecular perturbations obtained via normal mode sampling (NMS), and  
(iv) intermolecular interactions from Automated Interaction Site Screening (AISS).  

Together, these sources provide coverage of local chemical bonding, long-range dispersion, and relevant intramolecular degrees of freedom. Table~\ref{tab:dataset_hierarchy} summarizes the overall dataset composition.

\subsubsection*{Bayesian Optimization Structure Search (BOSS)}
BOSS~\cite{todorovic_bayesian_2019} was employed to explore adsorption geometries of molecules on Cu(111). A six-dimensional search space (two lateral translations, height, three rotations) was used with 20 initial points and 100 Bayesian iterations. Surface symmetries were exploited to reduce redundancy. Clash detection prevented unphysical overlaps, and an energy transformation down-weighted high-energy regions. Adsorption minima from the surrogate PES were refined with DFT and duplicates removed. The resulting minima were subsequently re-relaxed on Ag(111) and Au(111) to extend coverage without repeating the full search.

\subsubsection*{\emph{ab initio} molecular dynamics (AIMD)}
Born–Oppenheimer AIMD trajectories were generated for bulk Au, Ag, and Cu (3×3×3 and 5×5×5 supercells with 0–2 vacancies), surface slabs of the (111), (100), and (110) facets and selected molecule/metal systems. Frames were recorded every step and down-sampled by a factor of ten for MLIP training. Technical parameters are detailed in the SI.

\subsubsection*{Normal Mode Sampling}

To strengthen the description of the intramolecular degrees of freedom of the adsorbate molecules, we generated configurations by displacing atoms along their normal modes. Trajectories of the normal mode displacements were generated using the Atomic Simulation Environment (ASE)~\cite{ASE}, and the resulting structures were compared using the Smooth Overlap of Atomic Positions (SOAP) descriptor as implemented in the DScribe library~\cite{de_comparing_2016, himanen_dscribe_2020}, using a modified version of the code and methods reported in \cite{bhatia_leveraging_2025}. To avoid highly correlated configurations, we applied a farthest-point sampling algorithm to select a diverse subset. The filtered structures were then recomputed with DFT, yielding additional reference data that capture intramolecular degrees of freedom and subtle conformational changes relevant to adsorption.

\subsubsection*{Molecule–molecule interactions (AISS)}
The AISS module of \texttt{xtb}~\cite{aiss} was used to generate non-covalent dimers with GFN2-xTB~\cite{bannwarth_gfn2-xtbaccurate_2019}, spanning hydrogen bonding, $\pi$–$\pi$ stacking, dipole–dipole, halogen bonding, and dispersion interactions. Final single-point energies and forces were evaluated with FHI-aims.

\subsection*{MLIP training}
A typical training run for the MACE potentials was carried out on a single NVIDIA Tesla A100 GPU with a wall time of approximately 36 hours on Mahti (CSC IT Centre for Science). In terms of the models and typical training parameters, we employed message passing with 128 channels and max L = 1 (128x0e + 128x1o), 2 layers, each with correlation order 3, body order 4, and spherical harmonics up to l = 3. This implies 8 radial and 5 basis functions. Our radial cutoff was 5.0 Å, indicating a total receptive field of 10 Å for each atom. Hidden irreps were 128x0e + 128x1o, ADAM was used as the parameter optimizer with a typical batch size varying between 2 - 32 depending on the dataset and whether we were finetuning or training from scratch. The learning rate was 0.01 for all training from scratch, 0.0001 for the finetuning (multihead replay), weight decay 5$\times$$10^{-7}$, and the weighted energy and forces loss function was employed for all training. For accelerated training and inference, we employed the CuEquivariance library~\cite{cuEquivariance_nvidia}. To assess the effects of the training data composition, we partitioned the training set according to several filtering strategies or types. In the most comprehensive data sets, denoted \textit{full}, we include all data acquired from BOSS, NMS, AISS and relaxations, while for the data acquired through MD one snapshot is taken every ten frames due to the high correlation between individual frames. For the second type of training set, we further reduce the size of the training set by taking every one in five snapshots out of all types of configurations. This type of training set is termed \textit{small}. Following the initial filtering, we trained potentials with the \textit{full} and \textit{small} sets while varying the force/energy weight ratio ($\lambda$) between $\lambda = 1$ and $\lambda = 1000$, including the limiting cases of training only on forces or only on energies. Based on this, we selected appropriate $\lambda$ values for the subsequent training tests according to their performance. As a slightly more refined filtering method, we further employed the full MACE descriptors, filtering by their Euclidean distance using a number of thresholds (0.01–0.08) below which the configurations are considered too similar and subsequently discarded from the training set. For a computationally efficient similarity search suitable for a large number of high-dimensional vectors such as the MACE descriptors, we employed the GPU version of the Facebook AI Similarity Search (FAISS-gpu) tool. These datasets are denoted \textit{filtered} in the discussion. 

\subsection*{MLIP inference}
Relaxations and MD simulations on the training and test sets, molecular aggregates, biomolecules were implemented with python scripts leveraging ASE. For the larger simulations--organic monolayers and Au(111) reconstructions--we employed LAMMPS (10 Sep 2025 - Development version, Kokkos version 4.6.2~\cite{plimpton_fast_1995}) with the new unified ML-IAP interface, offering improved multi-GPU performance for both relaxations and dynamics. The MD simulations were done in the NVT ensemble with a 1.0 fs timestep.

\subsection*{Acknowledgements}

\subsection*{Author contributions}
M.G and J.S contributed equally to this work. 

\subsection*{Funding}
We acknowledge support from the Spanish Ministry of Science, Innovation and Universities, through project PID2023-149150OB-I00,  the predoctoral research contract PRE2021-098697, 
and the ``Mar\'{\i}a de Maeztu'' Programme for Units of Excellence in R\&D (CEX2023-001316-M) and Aalto Science Institute's Visiting Doctoral Researcher Programme. This work was supported by the World Premier International Research Centre Initiative (WPI), MEXT, Japan, and by the Research Council of Finland (Projects 347319 and 346824). The authors acknowledge the computational resources provided by the Aalto Science-IT Project and CSC IT Centre for Science, Espoo (Project 2008059).

\subsection*{Data Availability}

The code required to reproduce this work is freely available at \url{https://github.com/SINGROUP/MAD-SURF}. The data and models can be accessed through \url{https://zenodo.org/records/18312238}.

\bibliography{refs_clean}

\end{document}


\maketitle

\section{Model molecules and chemical motifs}
\begin{figure}[H]
    \centering
    \includegraphics[width=\linewidth]{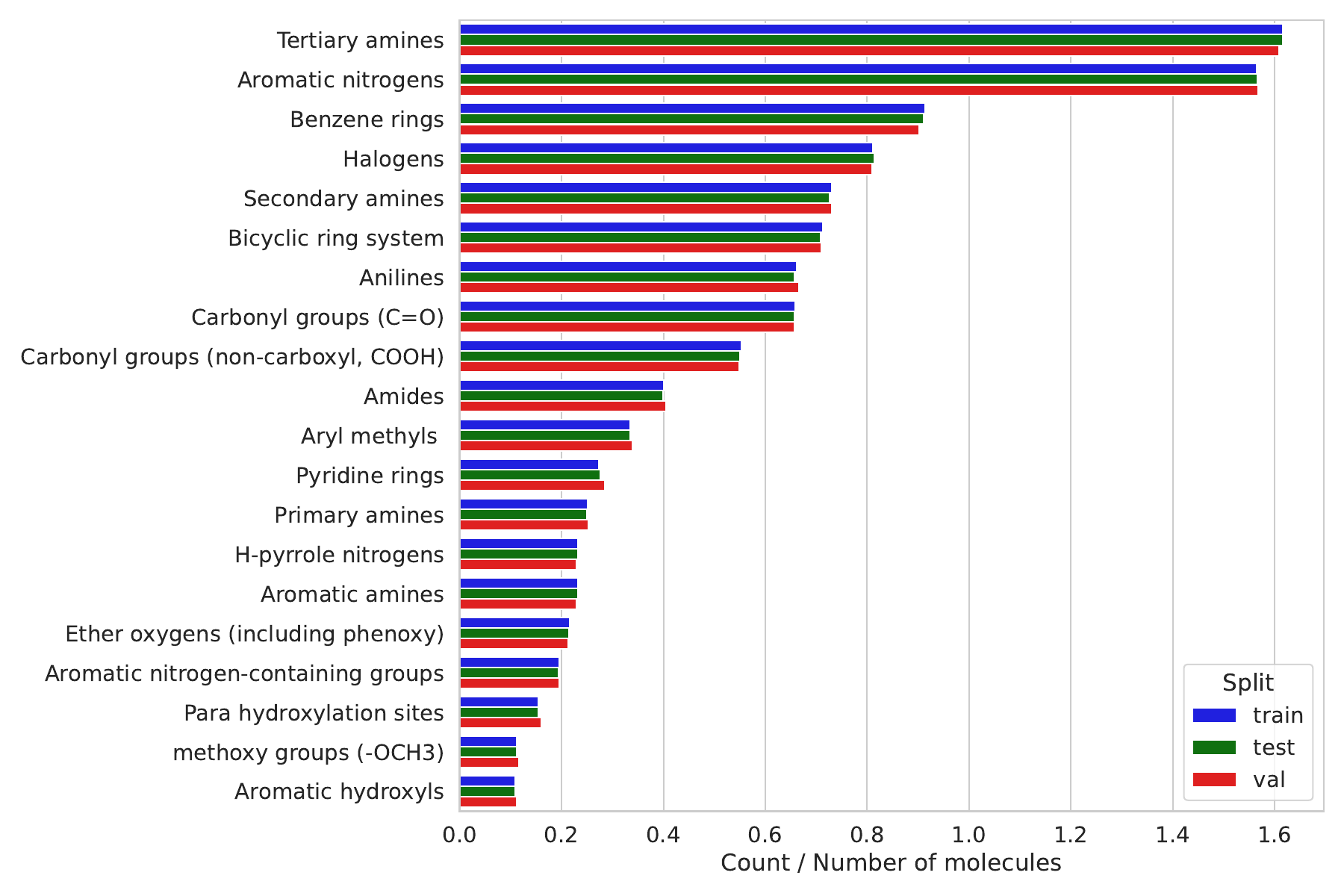}
\caption{Distribution of structural motifs in organic molecules relevant for scanning probe microscopy (SPM). 
The motif counts are based on our analysis of the QUAM-AFM database~\cite{GonzalezLastre2024Dec,QUAM-AFM_JCIM}, 
which contains 685,513 isolated quasi-planar molecules from PubChem. }

    \label{fig:motif_distribution}
\end{figure}

\begin{figure}[H]
    \centering
    \includegraphics[width=\linewidth]{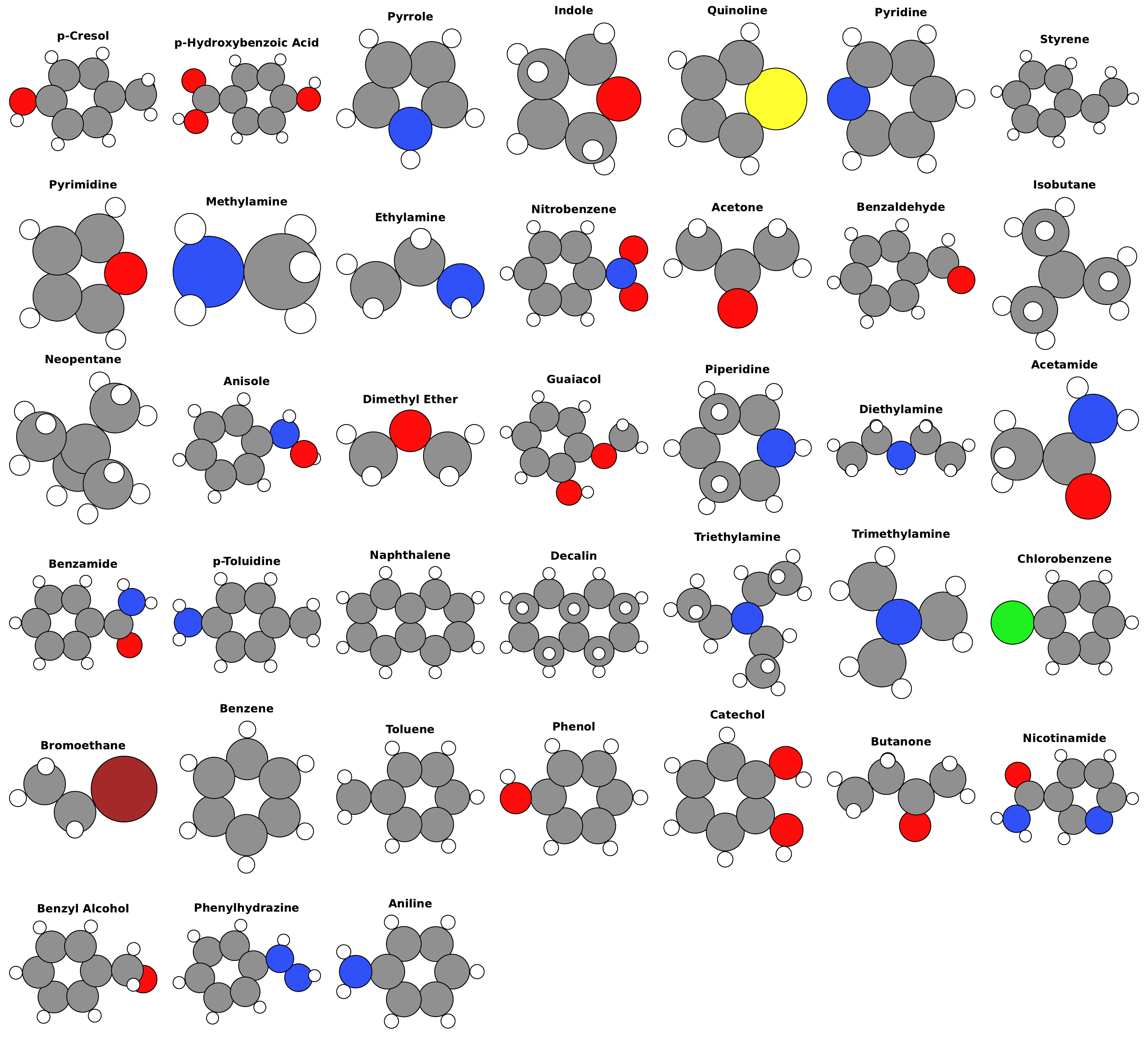}
\caption{Set of small molecules included in the training dataset. 
These molecules were selected to systematically cover the functional groups most relevant for SPM studies of organic adsorbates. 
Together they provide chemical building blocks that capture a wide spectrum of interactions with coinage metal surfaces, 
including strong N--metal coordination, hydrogen bonding, dispersion-dominated $\pi$-physisorption, and polarizable substituents.}

    \label{fig:small_mols}
\end{figure}

We selected molecules representing the functional groups most frequently encountered in SPM studies: $\pi$-systems, heteroatoms, hydrogen-bonding motifs, polar substituents, and carbonyl chemistry (Fig.~\ref{fig:motif_distribution}). 
Figure~\ref{fig:small_mols} illustrates motif coverage and selected molecules. Table~\ref{tab:pubchem_cids} lists all molecules with PubChem CIDs.
 Representative examples include:
\begin{itemize}
    \item \textbf{Aromatic systems}: benzene, toluene, naphthalene, pentacene, and phthalocyanine, as prototypical $\pi$-conjugated adsorbates.
    \item \textbf{Carbonyl-containing molecules}: acetone, benzaldehyde, and benzamide, to capture dipolar interactions and charge transfer.
    \item \textbf{Amines}: methylamine, diethylamine, aniline, and \textit{p}-toluidine, to model strong N--metal coordination common in on-surface chemistry.
    \item \textbf{Heteroaromatics}: pyridine, quinoline, and indole, probing nitrogen lone-pair interactions with metallic substrates.
    \item \textbf{Hydroxyl and methoxy groups}: phenol, catechol, guaiacol, and \textit{p}-cresol, introducing hydrogen bonding and surface anchoring.
    \item \textbf{Halogenated molecules}: chlorobenzene and bromoethane, exploring polarizable substituents relevant for molecular electronics and halogen bonding.
    \item \textbf{Benchmark SPM molecules}: perylene tetracarboxylic anhydride (PTCDA, archetypal planar aromatic with anhydride oxygens), triangulene precursors, 1,3,5-Tris(4-bromophenyl)benzene, Tetra(p-bromophenyl)porphyrin, pentacene, porphine and phthalocyanine (widely studied in AFM/STM and catalysis).
\end{itemize}

\begin{table}[ht]
\centering
\caption{Functional groups, representative molecules, and corresponding PubChem CIDs for training set molecules presented in Fig.~\ref{fig:small_mols}.}
\label{tab:pubchem_cids}
\small
\begin{tabular}{lll}
\toprule
\textbf{Functional group} & \textbf{Molecule} & \textbf{PubChem CID} \\
\midrule
Tertiary amines & Trimethylamine & 1146 \\
                & Triethylamine  & 8471 \\
\addlinespace

Aromatic nitrogens & Pyridine  & 1049 \\
                   & Quinoline & 8030 \\
\addlinespace

Benzene rings & Benzene & 241 \\
              & Toluene & 1140 \\
\addlinespace

Halogens & Chlorobenzene & 7964 \\
         & Bromoethane   & 6332 \\
\addlinespace

Secondary amines & Diethylamine & 8021 \\
                 & Piperidine   & 8082 \\
\addlinespace

Bicyclic ring systems & Decalin     & 7044 \\
                      & Naphthalene & 931 \\
\addlinespace

Anilines & \textit{p}-Toluidine & 7813 \\
\addlinespace

Carbonyl groups (C=O) & Acetone      & 180 \\
                      & Benzaldehyde & 240 \\
\addlinespace

Carbonyl groups (non-carboxyl, COOH) & Butanone & 6569 \\
\addlinespace

Amides & Acetamide & 178 \\
       & Benzamide & 2331 \\
\addlinespace

Aryl methyls & Benzyl alcohol & 244 \\
\addlinespace

Pyridine rings & Nicotinamide & 936 \\
\addlinespace

Primary amines & Methylamine & 6329 \\
               & Ethylamine  & 6341 \\
\addlinespace

H-pyrrole nitrogens & Pyrrole & 8027 \\
                    & Indole  & 8028 \\
\addlinespace

Aromatic amines & Aniline         & 6115 \\
                & Phenylhydrazine & 7516 \\
\addlinespace

Ether oxygens (including phenoxy) & Dimethyl ether & 8254 \\
                                  & Anisole        & 7518 \\
\addlinespace

Aromatic nitrogen-containing groups & Pyrimidine & 8029 \\
\addlinespace

Para hydroxylation sites & \textit{p}-Cresol              & 2879 \\
                         & \textit{p}-Hydroxybenzoic acid & 135 \\
\addlinespace

Methoxy groups & Guaiacol & 460 \\
\addlinespace

Aromatic hydroxyls & Phenol   & 996 \\
                   & Catechol & 289 \\
\addlinespace

Tertiary sp$^3$ hybridized carbon & Isobutane  & 6360 \\
                                  & Neopentane & 10041 \\
\addlinespace

Vinyl groups & Styrene & 7501 \\
\addlinespace

Nitro groups & Nitrobenzene & 7416 \\
\bottomrule
\end{tabular}
\end{table}

\section{Computational details}
For each molecule on a surface we leave 7 Å of distance between the molecule and its repeating image.

\subsection{BOSS setup}
Bayesian Optimization Structure Search (BOSS) was employed to explore molecular adsorption on Cu(111). The potential energy surface was sampled in a six-dimensional search space, consisting of two lateral translations ($x$, $y$), the adsorption height ($z$), and three rotations ($R_x, R_y, R_z$). Each search began with 20 random initial points followed by 100 Bayesian optimization iterations. Unphysical configurations were excluded by clash detection: if any atom–atom distance fell below 0.5 Å, the configuration was discarded and assigned a penalty energy of 5.0 eV. To reduce the weight of high-energy structures, energies above 1.0 eV were transformed to $1+\log(E)$~\cite{fang_efficient_2021}. The surrogate potential energy surface was then minimized to identify adsorption minima, which were further refined with full DFT relaxations. Duplicate minima converging to the same geometry were removed using the Kabsch algorithm~\cite{kabsch_solution_1976}. 

\subsection{AIMD setup}
We performed AIMD simulations for bulk and surface models of Au, Ag, and Cu, as well as for selected molecule/metal slab systems. Bulk systems were simulated using 3×3×3 and 5×5×5 supercells with 0–2 vacancies introduced to probe defect perturbations. Slab models of the (111), (100), and (110) facets were constructed as 5×5 lateral cells with three trilayers, where the bottom trilayer was kept fixed during dynamics. For molecule–surface systems, a minimum lateral separation of 7 Å between periodic images of adsorbates was imposed. Wavefunction extrapolation was disabled, and \texttt{MD\_restart} was set to \texttt{false} to avoid artificial correlations between consecutive MD steps. The $k$-point meshes employed for the different systems are summarized in Table~\ref{tab:aimd_kpoints}.

\begin{table}
\centering
\caption{Ab initio molecular dynamics (AIMD) configurations: system types, sizes, and $k$-point meshes used in FHI-aims. Each trajectory was run for 5~ps with a time step of 0.5~fs.}
\label{tab:aimd_kpoints}
\small
\begin{tabular}{lll}
\toprule
\textbf{System} & \textbf{Configuration} & \textbf{$k$-point grid} \\
\midrule
Bulk (Au, Ag, Cu) & $3\times3\times3$ supercell, 0–2 vacancies & $3\times3\times3$ \\
Bulk (Au, Ag, Cu) & $5\times5\times5$ supercell, 0–2 vacancies & $2\times2\times2$ \\
Surface (Au, Ag, Cu) & (111), $5\times5$ lateral, 3 trilayers & $2\times2\times1$ \\
Surface (Au, Ag, Cu) & (100), $5\times5$ lateral, 3 trilayers & $2\times2\times1$ \\
Surface (Au, Ag, Cu) & (110), $5\times5$ lateral, 3 trilayers & $2\times2\times1$ \\
Molecule/Surface & Adsorbates on above slabs, min. 7~\AA{} separation & same as bare slab \\
\bottomrule
\end{tabular}
\end{table}

\newpage

\section{Comparison of DFT and MAD-SURF results}

This section compares relaxed monolayer structures obtained with MAD-SURF against DFT reference calculations for representative adsorption systems. Superimposed displacement vectors quantify local deviations from the DFT geometries.

\begin{figure}[h!]
\begin{center}
\includegraphics[width=0.8\textwidth]{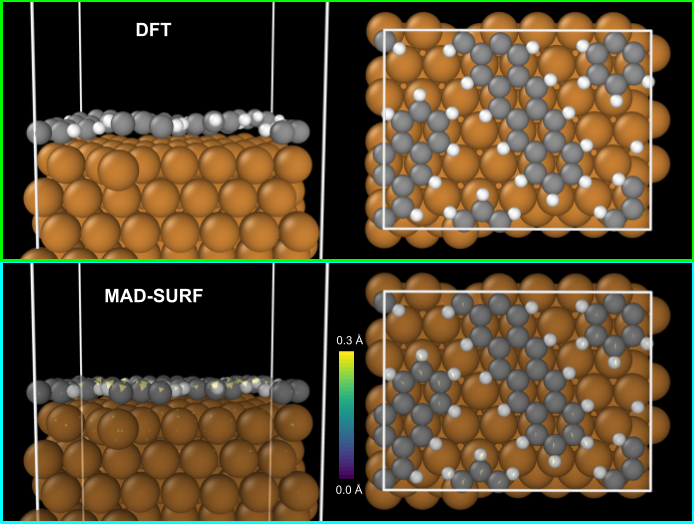}
\caption{
Comparison of the relaxed structural models for the HB monolayer of pentacene on Cu(111) obtained with the DFT reference method (top) and the MAD-SURF potential. 
}
\label{fig:monolayer_hb_pn_dft_vs_mlip}
\end{center}
\end{figure}

\begin{figure}[h!]
\begin{center}
\includegraphics[width=0.8\textwidth]{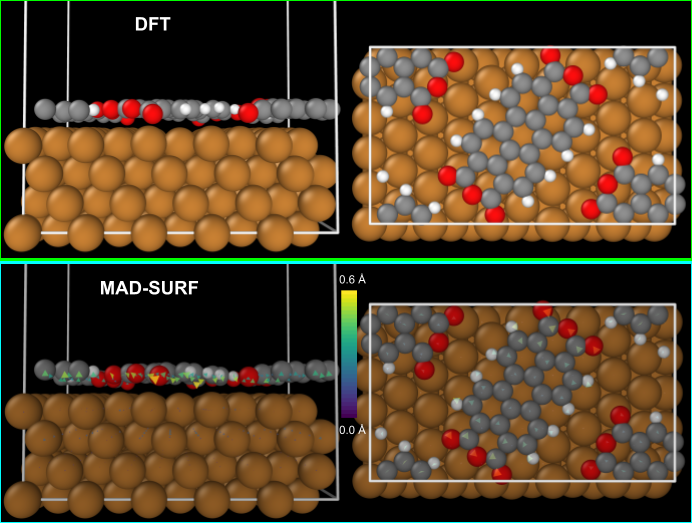}
\caption{
Comparison of the relaxed structural models for the HB monolayer of PTCDA on Cu(111) obtained with the DFT reference method (top) and the MAD-SURF potential. 
}
\label{fig:monolayer_hb_ptcda_cu111_dft_vs_mlip}
\end{center}
\end{figure}

\begin{figure}[H]
\begin{center}
\includegraphics[width=0.8\textwidth]{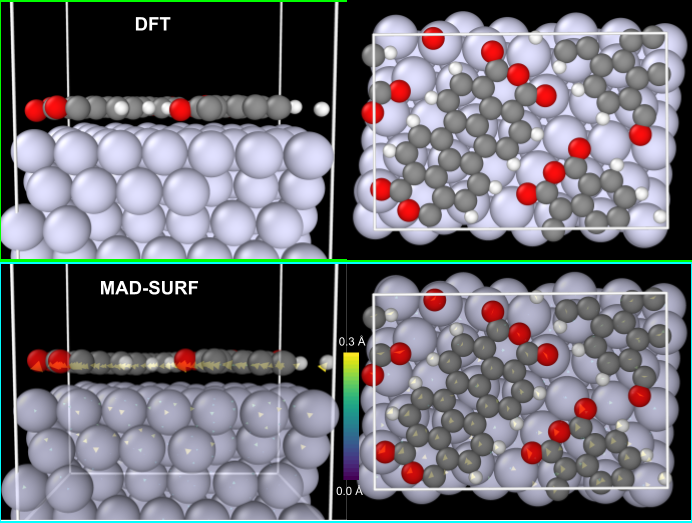}
\caption{
Comparison of the relaxed structural models for the HB monolayer of PTCDA on Ag(111) obtained with the DFT reference method (top) and the MAD-SURF potential. 
}
\label{fig:monolayer_hb_ptcda_ag111_dft_vs_mlip}
\end{center}
\end{figure}

\begin{figure}[H]
\begin{center}
\includegraphics[width=0.8\textwidth]{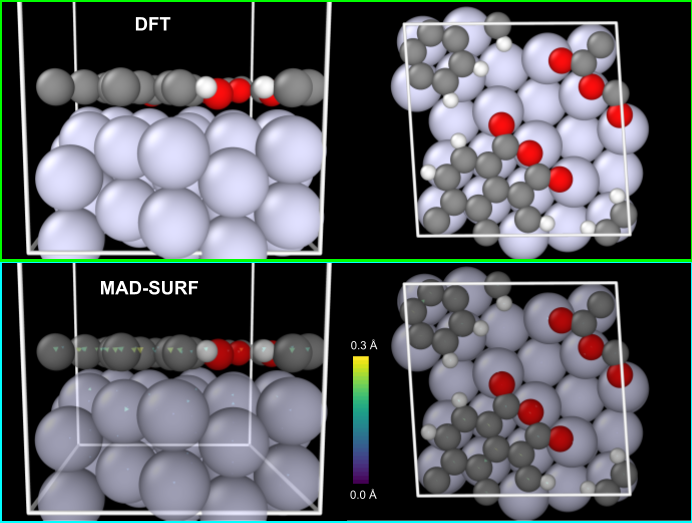}
\caption{
Comparison of the relaxed structural models for the BW monolayer of PTCDA on Ag(110) obtained with the DFT reference method (top) and the MAD-SURF potential. 
}
\label{fig:monolayer_bw_ptcda_dft_vs_mlip}
\end{center}
\end{figure}

\newpage

\section{$\beta$-Cyclodextrin OH orientation analysis}
\begin{figure}[b!]
    \centering
    \includegraphics[width=0.8\linewidth]{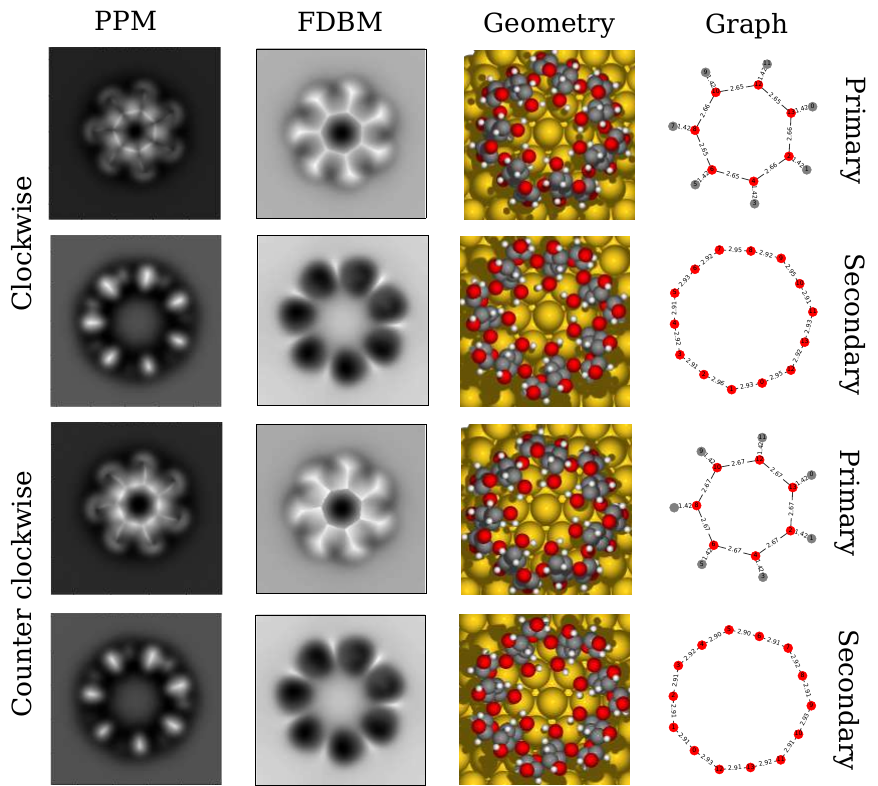}
    \caption{Probe Particle Model (PPM) and Full Density Based Model AFM simulations for $\beta$-CD/Au(111) comparing clockwise and counter clockwise OH network orientations. Shown are constant-height AFM images, the corresponding relaxed adsorption geometries, and graph representations of the OH network where edges are labeled by C-O and O–O distances for the primary and secondary faces.}
    \label{fig:beta_cd_graph}
\end{figure}

To assess whether the orientation of the OH network affects AFM contrast for this particular system, we compared clockwise and counter clockwise configurations of $\beta$-CD adsorbed on Au(111).
 For both the clockwise and counter clockwise orientation of the OH bonds, the AFM simulations (both PPM and FDBM) contrast is qualitatively indistinguishable for the primary and secondary face (Fig.~\ref{fig:beta_cd_graph}). No systematic changes in the position or intensity of the bright lobes are observed. A further analysis of the OH bonds and inter-O···O distance reveals that for both orientations the distance is very similar (1.42 \AA~of the C-O bonds and 2.66 \AA~for the inter-O···O), making these configurations indistinguishable from one another, consistent with previous analysis of hydrogen bond contrast in the AFM~\cite{ellner2019molecular, ellner2017}.

\clearpage

\bibliography{refs_clean}